\documentclass[aps,prl,twocolumn,groupedaddress,showpacs]{revtex4-1}

\usepackage{graphics}

\begin{document}

\title{Universal model for exoergic bimolecular reactions and inelastic processes}

% repeat the \author .. \affiliation  etc. as needed
% \email, \thanks, \homepage, \altaffiliation all apply to the current
% author. Explanatory text should go in the []'s, actual e-mail
% address or url should go in the {}'s for \email and \homepage.
% Please use the appropriate macro foreach each type of information

% \affiliation command applies to all authors since the last
% \affiliation command. The \affiliation command should follow the
% other information
% \affiliation can be followed by \email, \homepage, \thanks as well.
\author{Bo Gao}
\email[Email: ]{bo.gao@utoledo.edu}
%\homepage[Homepage: ]{http://bgaowww.physics.utoledo.edu}
%\thanks{}
\affiliation{Department of Physics and Astronomy,
	University of Toledo, Mailstop 111, 
	Toledo, Ohio 43606,
	USA}

%\date{\today}
\date{October 11, 2010}

\begin{abstract}

From a rigorous multichannel quantum-defect formulation of
bimolecular processes, we derive a fully quantal and analytic
model for the total rate of exoergic bimolecular reactions
and/or inelastic processes that is applicable over a wide range 
of temperatures including the ultracold regime. 
The theory establishes a connection
between the ultracold chemistry and the
regular chemistry by showing that the same theory
that gives the quantum threshold behavior 
agrees with the classical Gorin model
at higher temperatures. 
In between, it predicts that the rates
for identical bosonic molecules and distinguishable molecules
would first decrease with temperature outside of the Wigner
threshold region,
before rising after a minimum is reached.

\end{abstract}


\maketitle

The recent experiment by the JILA group \cite{osp10} represents
a milestone in studies of chemical reactions.
For the first time, reactions are studied in a temperature
regime where the quantum nature of the relative motion of
the reactants becomes unequivocally important,
as reflected in the quantum threshold behavior and 
in the importance of quantum statistics.
More importantly, the experiment strongly suggests that
bimolecular reactions in the ultracold regime follow universal
behaviors determined by the long-range interaction, 
as spelled out in more detail in related theoretical
works by Julienne and Idziaszek \cite{jul09,idz10} and 
by Qu\'em\'ener and Bohn \cite{que10}.

The significance of such experiments goes beyond
exploring chemical reactions in a new temperature regime 
with many unique characteristics, 
such as controllability via moderate
external fields \cite{Krems2008,ni10,que10}. 
By forcing a new perspective on the quantum
theory of reactions, as demanded by their interpretation, 
they have potential to improve our 
understanding of reactions and inelastic processes at all temperatures.
This paper is a illustration of such an outcome.
From a rigorous multichannel quantum-defect formulation
of bimolecular processes, we derive here a fully quantal model 
for the total rate of exoergic bimolecular reactions
and/or inelastic processes that is applicable over a wide range 
of temperatures. The theory establishes a connection
between the ultracold chemistry and the
regular chemistry by showing that the same theory
that gives the quantum threshold behavior \cite{osp10,jul09,idz10} 
agrees with the classical Gorin model \cite{gor38,fer06}
at higher temperatures. 
In between, it shows that the rates
for identical bosonic molecules and distinguishable molecules
would first decrease with temperature outside of the Wigner
threshold region,
before rising after a minimum is reached.
The theory further illustrates explicitly how the quantum
effects, including effects of quantum statistics,
gradually diminish at higher temperatures,
and establishes the van der Waals temperature scale as
the one that separates the quantum and the semiclassical behaviors
of reactions.
The same formalism is applicable to ion-molecule reactions
where our quantum model, with details to be presented 
in a separated publication, would approach the classical
Langevin model \cite{lan05,fer06} at high temperatures.

Consider the collision of two distinguishable molecules $A$ 
and $B$ in the absence of
any external fields. The cross section for a transition from
an entrance channel $i$ to an exit channel $f$ 
can be written in terms of the $S$ matrix as \cite{mot65}
\begin{eqnarray}
\sigma_{fi}(\epsilon) &=& \frac{\pi}{(2F_{Ai}+1)(2F_{Bi}+1)k_i^2} \nonumber\\
	& &\times\sum_{F_t,F_i,l_i,\{q_f\}} (2F_t+1)|S^{(F_t)}_{fi}-\delta_{fi}|^2 \;.
\end{eqnarray}
Here $\epsilon\equiv E-E_i=\hbar^2k_i^2/2\mu$ is the energy
relative to the entrance channel $i$, with $\mu$ being the
reduced mass.
$F_{A}$ and $F_{B}$ are the total (internal) angular
momenta of molecules $A$ and $B$, respectively.
$\mathbf{F}=\mathbf{F}_{A}+\mathbf{F}_{B}$ is the total angular momentum
excluding $l$, which is the relative angular momentum 
between $A$ and $B$.
$F_t$ is the total angular momentum of the system,
which is conserved in the absence of external fields.
$\{q_f\}$ represents the quantum numbers, excluding $F_t$,
that are required to characterize an exit channel $f$.

The exit channels can be classified into elastic channels,
labeled by $\{e\}$, inelastic channels, labeled by $\{u\}$, and
reactive channels, labels by $\{r\}$.
From the unitarity of the $S$ matrix \cite{mot65}, 
the total cross section, 
$\sigma_{\mathrm{ur}}\equiv\sum_{f\in\{u,r\}}\sigma_{fi}$, 
for the combination of all inelastic
and reactive processes, can be written as
\begin{eqnarray}
\sigma_{\mathrm{ur}}(\epsilon) &=& \frac{\pi}{(2F_{Ai}+1)(2F_{Bi}+1)k_i^2}\nonumber\\
	& &\times\sum_{F_t,F_i,l_i} (2F_t+1)(1-\sum_{F_e,l_e}|S^{(F_t)}_{ei}|^2) \;.
\label{eq:xsur}
\end{eqnarray}
The implication is that such a total cross section 
is completely determined by
the $S$ matrix elements for \textit{elastic channels only}.
The corresponding rate constant at temperature $T$ is
given in terms of $\sigma_{\mathrm{ur}}$ by 
\begin{equation}
K(T) = \left(\frac{8k_BT}{\pi\mu}\right)^{1/2}
	\frac{1}{(k_BT)^2}\int_0^\infty
	\epsilon\sigma_{\mathrm{ur}}(\epsilon)\exp(-\epsilon/k_BT)d\epsilon \;,
\label{eq:tave}	
\end{equation}
where $k_B$ is the Boltzmann constant.

Considerably further understanding of bimolecular
processes can be achieved through 
a multichannel quantum-defect theory (MQDT)
(see Ref.~\cite{gao05a} and references therein),
especially through an $S$ matrix formulation 
in terms of quantum reflection and transmission amplitudes associated with
the long-range potential \cite{gao08a}.
The theory, which is a multichannel generalization 
of the $S$ matrix formulation of Ref.~\cite{gao08a},
gives
\begin{eqnarray}
S^{(F_t)} &=& -(-1)^{l}\left[r^{(oi)}_{oo}
	+t^{(io)}_{oo}S^{c}_{\mathrm{eff}}
	(I-r^{(io)}_{oo}S^{c}_{\mathrm{eff}})^{-1}t^{(oi)}_{oo}\right] \;,
\label{eq:Smqdt1} \\
&=& -(-1)^{l}\left\{r^{(oi)}_{oo}
	+t^{(io)}_{oo}S^{c}_{\mathrm{eff}}
	\left[\sum_{m=0}^{\infty}(r^{(io)}_{oo}S^{c}_{\mathrm{eff}})^m
	\right]t^{(oi)}_{oo}\right\} \;.
\label{eq:Smqdt1e}
\end{eqnarray}
Here $(-1)^l$ is a diagonal matrix with elements $(-1)^{l_j}$
for channel $j$. The 
$r^{(oi)}_{oo}$ and $t^{(oi)}_{oo}$ are diagonal matrices
for the open channels with elements $r^{(oi)}_{l_j}(\epsilon_{sj})$
and $t^{(oi)}_{l_j}(\epsilon_{sj})$ representing the (complex)
quantum reflection
and the quantum transmission amplitudes, respectively, 
for molecules going outside-in (approaching each other) \cite{gao08a}.
They are universal functions of scaled energies,
$\epsilon_{sj}\equiv (E-E_j)/s_{Ej}$,
that are uniquely determined by the exponent, $\alpha_j$, of the
long-range interaction, $-C_{\alpha j}/R^{\alpha_j}_j$, in channel $j$,
and the $l_j$. Such long-range interactions have length scales
$\beta_{\alpha j}=(2\mu_j C_{\alpha j}/\hbar^2)^{1/(\alpha_j-2)}$ 
and corresponding energy scales 
$s_{Ej}=(\hbar^2/2\mu_j)(1/\beta_{\alpha j}^2)$, associated with them.
The $r^{(io)}_{oo}$ and $t^{(io)}_{oo}$ are similar, except that
their elements are amplitudes for molecules going inside-out (moving away from
each other) \cite{gao08a}. The $S^{c}_{\mathrm{eff}}$ is an effective short-range
$S$ matrix \cite{gao08a}, after the elimination of the closed 
channels \cite{gao05a}. It has the physical meaning of
being an effective reflection amplitudes by the 
inner potential.

Equation~(\ref{eq:Smqdt1}) for the $S$ matrix has a clear physical 
interpretation as discussed in Ref.~\cite{gao08a}.
In particular, the $m$-th term in its expansion,
Eq.~(\ref{eq:Smqdt1e}), corresponds to the contribution from
a path in which the fragments are reflected $m+1$ times
by the inner potential.
Further simplification can be achieved by
recognizing that $r^{(io)}_{l_j}(\epsilon_{sj})\approx 0$ for 
$\epsilon_{sj}\gg s_{Ej}$ \cite{gao08a}.
Dividing all open channels into elastic and near-degenerate
channels with $\epsilon_{sj}<\sim s_{Ej}$, and
other channels with $\epsilon_{sj}\gg s_{Ej}$,
we have
\begin{eqnarray}
S^{(F_t)}_{ei} &\approx& -(-1)^{l_e}\left\{r^{(oi)}_{l_i}
	\delta_{ei}\right.\nonumber\\
	& &+\left.t^{(io)}_{l_e}\left[\widetilde{S}^{c}_{\mathrm{eff}}
	(1-r^{(io)}\widetilde{S}^{c}_{\mathrm{eff}})^{-1}\right]_{ei}t^{(oi)}_{l_i}\right\} \;,
\label{eq:Smqdt2}
\end{eqnarray}
where $\widetilde{S}^{c}_{\mathrm{eff}}$ is a
submatrix of the effective short-range 
$S^{c}_{\mathrm{eff}}$ that includes only the elastic
and other near-degenerate channels for which 
the quantum reflection amplitude $r^{(io)}_l$ 
differs substantially from zero.

A number of different theories and models, both exact and approximate, 
can be derived from 
Eq.~(\ref{eq:xsur}), and either Eq.~(\ref{eq:Smqdt1}) or (\ref{eq:Smqdt2}).
The universal model to be presented here,
which we call the quantum Langevin (QL) model, 
results from the assumption 
of no reflection by the inner potential, namely,
\begin{equation}
\widetilde{S}^{c}_{\mathrm{eff}}\approx 0 \;.
\label{eq:LangAssump}
\end{equation}
It is a rigorous mathematical
representation of the Langevin assumption \cite{lan05,fer06}
in a quantum theory.
In plain language, it assumes that whenever two molecules
come sufficiently close
to each other, so many ``bad'' things can happen that they
can never get out of it in their initial configurations.
It can be expected to be a good approximation whenever there are
a large number of open exit channels that are strongly coupled to
the entrance channel in the inner region. For it 
to be satisfied in the limit of zero energy, the reactions and
inelastic processes under consideration have to be at least exoergic.

Under the Langevin assumption, Eq.~(\ref{eq:Smqdt2}) gives
\begin{equation}
S^{(F_t)}_{ei} \approx -(-1)^{l_i}r^{(oi)}_{l_i}\delta_{ei}\;.
\label{eq:SQT}
\end{equation}
It implies that the elastic $S$ matrix elements in 
the QL model, and therefore
the total cross section and the corresponding total rate
for inelastic and reactive processes, are all described
by universal functions that are uniquely determined by
the long-range interaction in the entrance channel.
Substituting Eq.~(\ref{eq:SQT}) 
into Eq.~(\ref{eq:xsur}) and subsequently into Eq.~(\ref{eq:tave}),
the total rate of inelastic and reactive processes
in the QL model can be written as
\begin{equation}
K(T) = s_K {\cal K}^{(\alpha)}(T_s) \;.
\end{equation}
Here $s_K$ is the rate scale corresponding to the long-range,
$-C_\alpha/R^\alpha$, interaction in the entrance channel. 
\begin{equation}
s_K = (\hbar/\mu\beta_{\alpha})\pi\beta_{\alpha}^2 = \pi\hbar\beta_{\alpha}/\mu \;,
\end{equation}
in which $\hbar/\mu\beta_{\alpha}$ is the velocity scale corresponding
to the length scale $\beta_{\alpha}$.
${\cal K}^{(\alpha)}(T_s)$ is a universal function of the
scaled temperature, $T_s = T/(s_E/k_B)$, that is uniquely
determined by the exponent $\alpha$. Specifically,
\begin{equation}
{\cal K}^{(\alpha)}(T_s) = \frac{2}{\sqrt{\pi}}
	\int_0^\infty dx\: x^{1/2} e^{-x}
	{\cal W}^{(\alpha)}(T_s x)\;,
\end{equation}
where ${\cal W}^{(\alpha)}(\epsilon_s)$ is a scaled total rate
before thermal averaging. It depends on energy only through the
scaled energy $\epsilon_s = \epsilon/s_E$, and has contributions
from all partial waves:
\begin{equation}
{\cal W}^{(\alpha)}(\epsilon_s) = \sum_l{\cal W}^{(\alpha)}_{l}(\epsilon_s) \;.
\end{equation}
Here ${\cal W}^{(\alpha)}_{l}$ is a scaled partial rate
given by
\begin{equation}
{\cal W}^{(\alpha)}_{l}(\epsilon_s)
	=(2l+1){\cal T}^{c(\alpha)}_l(\epsilon_s)/\epsilon_s^{1/2} \;,
\end{equation}
in which ${\cal T}^{c(\alpha)}_l(\epsilon_s)=|t^{(oi)}_l(\epsilon_s)|^2$
is the quantum transmission probability through the long-range
potential at the scaled energy $\epsilon_s$ and for partial wave $l$
\cite{gao08a}.

This QL model for reactions and inelastic processes is applicable to both
neutral-neutral systems, for which $\alpha=6$ corresponding to the
van der Waals potential, and charge-neutral
systems, for which $\alpha=4$ corresponding to the polarization
potential. 
We focus here on the neutral-neutral case
to make connection with existing theories and experiments \cite{osp10,jul09,idz10,que10}.
The results for charge-neutral systems will be presented elsewhere.

\begin{figure}[ht]
\scalebox{0.4}{\includegraphics{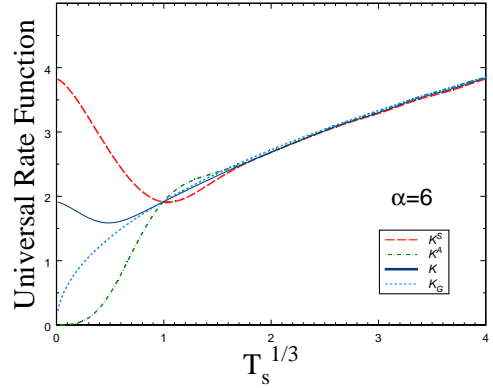}}
\caption{(Color online) The universal rate functions
${\cal K}^{S(\alpha)}(T_s)$, ${\cal K}^{A(\alpha)}(T_s)$, and
${\cal K}^{(\alpha)}(T_s)$ for $\alpha=6$, corresponding to 
$-1/R^6$ type of interaction 
in the entrance channel.
Here ${\cal K}_G$ refers to the prediction of the classical
Gorin model, as given by Eq.~(\ref{eq:Gorin}). All results
include summations over all relevant partial waves.
\label{fig:urf6}}
\end{figure}

For $\alpha=6$, the quantum transmission probability through
the long-range potential, ${\cal T}^{c(\alpha)}_l(\epsilon_s)$, which
is the only quantity required to determine the universal
rate functions in the QL model, can be found analytically
by substituting Eqs.~(A1)-(A4)
of Ref.~\cite{gao08a} into the Eq.~(52) of the same reference. 
The result is
\begin{equation}
{\mathcal T}^{c(6)}_l(\epsilon_s) = \frac{2M_{\epsilon_s l}[\cos(\pi\nu)-\cos(3\pi\nu)]}
	{1-2M_{\epsilon_s l}\cos(3\pi\nu)+M_{\epsilon_s l}^2}\;.
\label{eq:tp6}	
\end{equation}
Here $\nu$ is the characteristic exponent for 
$-1/R^6$ potential \cite{gao98a}, and
\begin{eqnarray}
M_{\epsilon_s l}(\nu) &=& |\Delta|^{2\nu}\left[\frac{\Gamma(1-\nu)}{\Gamma(1+\nu)}\right]
	\left[\frac{\Gamma(1+\nu_0-\nu)}{\Gamma(1+\nu_0+\nu)}\right] \nonumber\\
	& &\times \left[\frac{\Gamma(1-\nu_0-\nu)}{\Gamma(1-\nu_0+\nu)}\right]
	\left[\frac{C_{\epsilon_s l}(-\nu)}{C_{\epsilon_s l}(\nu)}\right] \;,
\label{eq:Mnu}
\end{eqnarray}
where $\Delta = \epsilon_s/16$, $\nu_0 = (2l+1)/4$, and
\begin{equation}
C_{\epsilon_s l}(\nu) = \prod_{j=0}^{\infty} Q(\nu+j) \;,
\label{eq:cj}
\end{equation}
in which $Q(\nu)$ is given by a continued 
fraction:
\begin{equation}
Q(\nu) = \frac{1}{1-\Delta^2\frac{1}{(\nu+1)
	[(\nu+1)^2-\nu_0^2](\nu+2)[(\nu+2)^2-\nu_0^2]}
	Q(\nu+1)} \;.
\label{eq:Qcf}
\end{equation}
The resulting universal rate function, ${\cal K}^{(\alpha)}(T_s)$,
applicable to neutral-neutral distinguishable molecules, is illustrated in
Figure~\ref{fig:urf6}.
Similar results can be obtained for neutral-neutral interactions 
of identical molecules, following considerations similar to those
of Ref.~\cite{gao96}. 
They are given generally by a combination of two
universal rate functions defined by
\begin{equation}
{\cal K}^{S(\alpha)}(T_s) = \frac{2}{\sqrt{\pi}}
	\int_0^\infty dx\: x^{1/2} e^{-x}
	{\cal W}^{S(\alpha)}(T_s x)\;,
\end{equation}
where
\begin{equation}
{\cal W}^{S(\alpha)}(\epsilon_s) = 2\sum_{l=\mathrm{even}}{\cal W}^{(\alpha)}_{l}(\epsilon_s) \;,
\end{equation}
and
\begin{equation}
{\cal K}^{A(\alpha)}(T_s) = \frac{2}{\sqrt{\pi}}
	\int_0^\infty dx\: x^{1/2} e^{-x}
	{\cal W}^{A(\alpha)}(T_s x)\;,
\end{equation}
where
\begin{equation}
{\cal W}^{A(\alpha)}(\epsilon_s) = 2\sum_{l=\mathrm{odd}}{\cal W}^{(\alpha)}_{l}(\epsilon_s) \;.
\end{equation}
For example, in terms of ${\cal K}^{S(\alpha)}$ and ${\cal K}^{A(\alpha)}$,
$K^S(T)=s_K{\cal K}^{S(\alpha)}$ gives the rate
for identical bosonic molecules in the same internal ($M$) state, and 
$K^A(T)=s_K{\cal K}^{A(\alpha)}$ gives the rate
for identical fermionic molecules in the same internal state.
The three rate functions are related by
${\cal K}^{(\alpha)}=({\cal K}^{S(\alpha)}+{\cal K}^{A(\alpha)})/2$,
and are all illustrated in Figure~\ref{fig:urf6} for $\alpha=6$.

At ultracold temperatures such that $T_s \ll 1$, a 
QDT expansion \cite{gao09a}
of ${\mathcal T}^{c(6)}_l(\epsilon_s)$ gives
\begin{equation}
{\cal K}^{S(6)}(T_s) = 8\bar{a}_{sl=0}\left[
	1-\frac{4\bar{a}_{sl=0}}{\sqrt{\pi}}T_s^{1/2}
	+3\bar{a}_{sl=0}^2T_s + O(T_s^{3/2})\right]\;,
\label{eq:KSexp}	
\end{equation}
where $\bar{a}_{sl=0} = 2\pi/[\Gamma(1/4)]^{2} \approx 0.4779888$
is the scaled mean scattering length for $l=0$ \cite{gao09a},
\begin{equation}
{\cal K}^{A(6)}(T_s) = 36\bar{a}_{sl=1}T_s\left[
	1-\frac{16\bar{a}_{sl=1}}{\sqrt{\pi}}T_s^{3/2}
	+ O(T_s^{2})\right]\;,
\label{eq:KAexp}	
\end{equation}
where $\bar{a}_{sl=1} = [\Gamma(1/4)]^{2}/36\pi \approx 0.1162277$
is the scaled mean scattering length for $l=1$ \cite{gao09a},
and
\begin{eqnarray}
{\cal K}^{(6)}(T_s) &=& 4\bar{a}_{sl=0}
	-\frac{(4\bar{a}_{sl=0})^2}{\sqrt{\pi}}T_s^{1/2} \nonumber\\
	& &+\left(12\bar{a}_{sl=0}^3+18\bar{a}_{sl=1}\right)T_s 
	+ O(T_s^{3/2})\;.
\label{eq:Kexp}	
\end{eqnarray}
At high temperatures as characterized by $T_s\gg 1$, 
it is straightforward to show, from the semiclassical limit
of the transmission probabilities \cite{gao08a}, that
\begin{equation}
{\cal K}^{S(6)}(T_s)\approx{\cal K}^{A(6)}(T_s)
\approx {\cal K}^{(6)}(T_s)\sim \frac{2^{4/3}\Gamma(2/3)}{\sqrt{\pi}}T_s^{1/6} \;,
\label{eq:Gorin}
\end{equation}
in agreement with the classical Gorin model \cite{gor38,fer06}.
All scaled results can be put on absolute scales using a
single parameter, the $C_6$ coefficient for the entrance channel,
from which both the temperature scale $s_E/k_B$ and the rate
scale $s_K$ can be determined \cite{sup}.

%\textit{Predictions and Implications}:
In the Wigner threshold region, in which the rates are accurately
characterized by the first terms of Eqs.~(\ref{eq:KSexp})-(\ref{eq:Kexp}),
our results are consistent with those of Julienne and Idziaszek \cite{jul09,idz10}.
Outside of this region,
both ${\cal K}^{(6)}$ and ${\cal K}^{S(6)}$ 
are predicted to first decrease with
temperature, a behavior that
deviates strongly from the prediction of the classical
Gorin model. Specifically, ${\cal K}^{(6)}$  is predicted to reach a minimum
value of ${\cal K}^{(6)}_{\mathrm{min}}\approx 1.587$ at
$T_{s\mathrm{min}}^{(6)}\approx 0.1154$, for a drop of about
17\% from its value at zero temperature.
The ${\cal K}^{S(6)}$ is predicted to reach a minimum
value of ${\cal K}^{S(6)}_{\mathrm{min}}\approx 1.908$ at
$T^{S(6)}_{s\mathrm{min}}\approx 1.114$, for a drop of about
50\% from its value at zero temperature.
For the JILA experiment \cite{osp10},  
$T_{s\mathrm{min}}^{(6)}$ translates,
using the $C_6$ coefficients of Kotochigova \cite{kot10,sup},
to $T_{\mathrm{min}}^{(6)}\approx 2.58$ $\mu$K for 
$^{40}$K$^{87}$Rb+$^{40}$K$^{87}$Rb
in different internal states, and to 
$T_{\mathrm{min}}^{(6)}\approx 11.9$ $\mu$K for 
$^{40}$K+$^{40}$K$^{87}$Rb.
It is worth noting that an experimental measurement of either 
$T_{\mathrm{min}}^{(6)}$ or $T^{S(6)}_{\mathrm{min}}$
would constitute a measurement of the $C_6$ coefficient,
a fact that can be valuable especially for more complex molecules
for which theoretical calculations of $C_6$ \cite{der99,kot10} become
increasingly difficult and unreliable.
At higher temperatures, our results show how the 
quantum effects, including that of quantum statistics, 
gradually diminish, and all rates approach that 
of the classical Gorin model \cite{gor38,fer06}.
As illustrated in Fig.~\ref{fig:urf6}, such a transition
from quantum to semiclassical behavior occurs over
a range of the van der Waals temperature scale $s_E/k_B$.

The QL model gives the total rate that includes both reactive and
inelastic processes. For experiments with only
reactive channels open \cite{osp10}, it give the total rate
of reactions. For experiments with no open reactive channels,
it gives the total rate of inelastic processes.
In all case, the requirement for its validity 
is that there are many open channels
that are strongly coupled to the entrance channel
by the short-range interactions. Of interest 
in the context of cold-atom physics, the QL model serves 
to unify theories of
ultracold chemistry \cite{jul09,idz10,que10,kot10} with theories for 
atom-atom \cite{orz99}, atom-molecule \cite{hud08}
and molecule-molecule inelastic processes.
For example, for a vibrational highly excited molecule,
except Feshbach molecules with with very small 
binding energies \cite{kno10},
the theory predicts that its collisional lifetime is approximately
independent of its initial state, and that the rates 
for atom-molecule and molecule-molecule 
inelastic processes are related.
More specifically, for Cs$_2$ in a highly excited
rovibrational state \cite{dan08},
it predicts that the  Cs$_2$-Cs$_2$ inelastic
rate should have a minimum around 6.31 $\mu$K
(assuming that they are prepared in the same state),
and Cs$_2$-Cs inelastic rate has a minimum around
1.70 $\mu$K. More detailed discussion of
such applications will be presented elsewhere.

%\textit{Conclusions}: 
In conclusion, we have presented a universal model of 
exoergic bimolecular reactions and/or inelastic processes
that is applicable over a wide range of temperatures,
illustrating the evolution from quantum behavior to
semiclassical behavior. It is an important baseline
model in which rates for different systems differ from each other
only in scaling, and has an intriguing and useful property
of being more accurate for more complex systems. 
Simple analytic formulas, to higher orders
than those of Julienne and Idziaszek \cite{jul09,idz10}, 
and applicable over a substantially wider range of temperatures, 
are also presented. 
Equally important, we believe, is that the underlying MQDT formulation,
such as Eq.~(\ref{eq:Smqdt2}), lays a solid foundation for new types of 
theories of reactions and inelastic processes, either rigorous or 
approximate, that goes further beyond the QL model.

\begin{acknowledgments}
I thank Jun Ye and Timur Tscherbul for motivations and helpful discussions.
This work was supported by NSF under the Grant number PHY-0758042.
\end{acknowledgments}

\bibliography{bgao,twobody,chem}

\end{document}